\documentclass[fleqn,usenatbib]{mnras}

\usepackage{newtxtext,newtxmath}

\usepackage[T1]{fontenc}

\DeclareRobustCommand{\VAN}[3]{#2}
\let\VANthebibliography\thebibliography
\def\thebibliography{\DeclareRobustCommand{\VAN}[3]{##3}\VANthebibliography}

\usepackage{graphicx}	
\usepackage{amsmath}	
\usepackage{float}

\title[Exotic radio source discovery with VQ-VAEs]{Prospecting MeerKAT Continuum Data for Enigmatic Radio Sources with Unsupervised Vector-Quantised Variational Autoencoders}

\author[Ventura, Thorat, Bosman, Deane and Cleghorn]{
Fernando L. Ventura,$^{1,2}$
Kshitij Thorat,$^{1}$
Anna Bosman,$^{3,4}$
Roger Deane,$^{5,6,1}$ \thanks{E-mail: roger.deane@wits.ac.za}
Christopher Cleghorn$^{7}$
\\
$^{1}$Department of Physics, University of Pretoria, Private Bag X20, Pretoria 0028, South Africa \\
$^{2}$Inter-University Institute for Data Intensive Astronomy\\
$^{3}$Department of Computer Science, University of Pretoria, Private Bag X20, Pretoria 0028, South Africa  \\
$^{4}$African Institute for Data Science and Artificial Intelligence (AfriDSAI), University of Pretoria, Private Bag X20, Pretoria 0028, South Africa \\
$^{5}$Wits Centre for Astrophysics, School of Physics, University of the Witwatersrand, 1 Jan Smuts Avenue, Johannesburg, 2000, South
Africa \\
$^{6}$Inter-University Institute for Data Intensive Astronomy, Department of Astronomy, University of Cape Town, Cape Town, South Africa \\
$^{7}$School of Computer Science and Applied Mathematics, University of the Witwatersrand,
Jan Smuts Ave, Johannesburg, South Africa\\
}

\pubyear{2026}

\begin{document}
\label{firstpage}
\pagerange{\pageref{firstpage}--\pageref{lastpage}}
\maketitle

\begin{abstract}

 We present a novel application of Vector quantised variational autoencoders (VQ-VAEs) as an unsupervised tool to probe deep 1.28 GHz radio continuum images taken from the MeerKAT Galaxy Cluster Legacy Survey (MGCLS) and compare their performance to other machine learning methods. We examine the effectiveness of VQ-VAEs in identifying radio continuum sources with anomalous structures in the image-plane domain. We compare performance to a supervised training set and other published anomaly detection methods, focusing especially on simple autoencoders (AE), Memory Unit Autoencoders (MUAE) and a human-in-the-loop software \texttt{Astronomaly} based on bootstrap your own latent (BYOL). Our investigations show that VQ-VAE perform better than simple AEs, are not as accurate as MUAE but are much faster and are also faster and comparable to \texttt{Astronomaly} while not requiring labelled data. We observe that they are able to remove a majority of the typical sources in such data, even when trained in an unsupervised manner on unlabelled data. We also provide our testing set of a large sample of manually labelled radio sources, in particular radio galaxies, taken from the MGCLS at 1.28 GHz. We find VQ-VAEs to be potentially useful as part of automated approaches towards searching through high volumes of data which are key in extracting the full scientific potential of the Square Kilometre Array and its pathfinders.

\end{abstract}




\section{Introduction}

Although machine learning (ML) is still relatively young in radio astronomy, it is expected to play a central role in addressing the field’s growing big-data challenges. A particularly relevant problem is the automated identification of exotic or anomalous sources in increasingly deep, wide, high-fidelity radio imaging surveys. Such anomalies are often of great scientific interest, as they may point to previously unrecognised astrophysical processes or rare transient phenomena. However, the definition of what constitutes an anomaly can be ambiguous and subjective, further complicated by the fact that physically unusual sources may not be visually distinct in imaging data.

Unsupervised and semi-supervised ML methods, such as autoencoders (AEs), offer a promising avenue for anomaly detection in astronomy due to their ability to learn compact representations of data without requiring extensive labelled datasets~\citep{han2022identifying, alonso2024detecting, Brand2026}. AEs consist of two parts, an encoder model and a decoder model. The encoder compresses the input data into a low-dimensional latent space and the decoder subsequently reconstructs the input data. The reconstruction error can serve as a proxy for novelty. In this approach, simple, common sources are easily reconstructed with low residual error. 

In convolutional AEs, the encoder and decoder are convolutional neural networks (CNNs). CNNs are multi-layered neural networks designed to learn spatial correlations between neighbouring input signals. As an input image is passed through the model, the input for each layer is convolved by some kernel to produce the output which serves as the input for the next layer. The kernel weights are iteratively adjusted during training to minimise the reconstruction error.

The application of ML techniques for anomaly detection in astronomy has gained significant attention, particularly with the advent of large-scale surveys that produce vast amounts of data~\citep{Lochner2021, giger2024unsupervised, Protege}. Traditional approaches, such as AEs and VAEs, have been employed to model the distribution of ``normal'' data, and identify deviations indicative of anomalies~\citep{Doorenbos2021,andrianomena2024radio,Brand2026}. \citet{Doorenbos2021} compared AEs, including convolutional AEs, to four other ML approaches for use in anomaly detection to outlier detection on the Sloan Digital Sky Survey (SDSS). 
They found that when comparing a suite of algorithms, AEs selected the most distinct sample of outliers. \citet{Brand2026} performed further comparisons between convolutional AEs and traditional ML techniques for the purpose of anomaly detection in radio astronomical data, and concluded that AEs outperformed conventional ML when trained on independent principal components extracted from the data.

\begin{figure}
\centering
\includegraphics[width=0.5\textwidth]{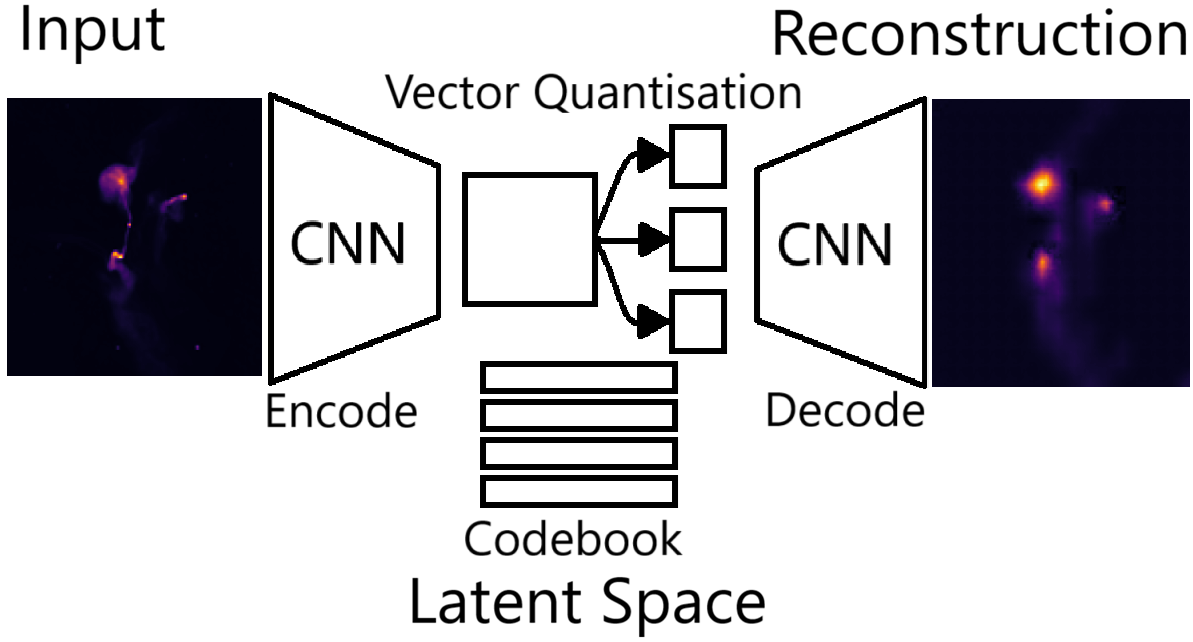}
\caption[Diagram of the basic structure of a VQ-VAE]{Basic structure of a VQ-VAE. The encoder reduces the input to the dimensions of the latent space. A code form is produced from a learned codebook of discrete vectors. The decoder attempts to reconstruct the original input from the code form.}
\label{fig:autoencdia}
\end{figure}

Variational autoencoders (VAEs) and their discrete counterparts, vector-quantised VAEs (VQ-VAEs) \citep{vqvae}, improve upon classical AEs by learning structured latent representations that are more amenable to clustering and generative tasks~\citep{van2017neural}. VQ-VAEs in particular, schematically illustrated in Figure~\ref{fig:autoencdia}, are capable of learning a discrete codebook of latent embeddings, making them well-suited to anomaly detection, where clear distinctions between common and rare patterns are desirable~\citep{marimont2021anomaly,vqvae_AD}. VQ-VAEs offer several key advantages over standard AEs and VAEs for anomaly detection~\citep{marimont2021anomaly,vqvae_AD}, particularly in domains like astronomy, where the structure and diversity of the data are high, and anomalous sources may only be subtly different. These advantages stem primarily from the nature of latent space representation of VQ-VAEs, which is discrete and quantised, as opposed to continuous and often entangled in standard AEs and VAEs~\citep{van2017neural}. In the medical imaging domain, VQ-VAEs have been successfully applied to detect anomalies in optical coherence tomography angiography (OCTA) images, demonstrating their potential to identify subtle pathological features~\citep{jebril2024anomaly}. Similarly, \citet{pinaya2022unsupervised} combined VQ-VAEs with an ensemble of Transformer networks to detect anomalies in 3D brain imaging. These studies underscore the potential of VQ-VAEs to capture complex data distributions and detect anomalies in real-world domains.

In this paper, we explore the feasibility of using VQ-VAEs as an off-the-shelf architecture for unsupervised anomaly detection in radio astronomical data. We use the MeerKAT Galaxy Cluster Legacy Survey \citep[MGCLS,][]{Knowles2022}, from which we create a catalogue of labelled sources, as well as many unlabelled sources as identified by source-finding software. The labelled sources are manually labelled into typical and exotic sources. Although the definition of an exotic source is subjective, we attach the label to those sources which we believe are scientifically interesting due to their scarcity or how unusual they may be. For instance, we label X-shaped sources as exotic, while bent sources, common FRIs, and FRIIs are labelled typical. A physically anomalous source may well have a visually typical appearance. We investigate how well a convolutional VQ-VAE can find exotic sources in the data. We compare the performance of VQ-VAEs trained on the unlabelled data to other published machine learning tools in astronomy and evaluate the particular weaknesses and strengths of the model and where the model may be useful in comparison.

\section{Data}

To test the utility of VQ-VAE's specifically for SKA precursor data, we chose the publicly available MeerKAT Galaxy Cluster Legacy Survey (MGCLS) Stokes-I images. MGCLS is formed of deep MeerKAT images of 115 galaxy clusters with high fidelity in the L-band (900-1670 MHz). Each cluster field was observed for about 6 to 10 hours in full polarisation mode.  Each image has a central observing frequency of 1.28~GHz with typical sensitivity ranging from about 3 to 5~\textmu{}Jy/beam and imaging dynamic range of order 10$^{4-5}$. The images have a typical angular resolution of \textasciitilde~10~arcsec and high sensitivity to structures up to about 10~arcmin scales. The basic full-field spectral cubes span about 2~deg\(\times\)2~deg while the enhanced products consist of the inner 1.2~deg by 1.2~deg field of view corrected for the primary beam. A wide field of view means there are hundreds to thousands of non-cluster sources per image. The compact source catalogue, presented in \cite{Knowles2022}, contains approximately 626,000 sources, not counting the more challenging extended, diffuse sources whose environments are "contaminated" with these often unrelated compact / point sources. It should be noted that these images as such are fundamentally different from a perspective of spatial source density and complexity from typical samples which use samples from older surveys such as Faint Images of Radio SKy at Twenty Centimeter (FIRST) and NRAO-VLA Sky Survey (NVSS) (see \citep{Becker1995} and \citep{Condon1998} respectively), and therefore closer to the images which we may expect from the upcoming SKA. 

The galaxy cluster images are processed to find extended sources using the source-finding software \texttt{PyBDSF}\footnote{\url{https://github.com/lofar-astron/PyBDSF}}. \texttt{PyBDSF} searches for and models sources in the larger image using ellipses, which it groups into islands of emission which may represent an extended source. The thresholds for included emission and grouping are adjusted by trial and error until \texttt{PyBDSF} produces a catalogue containing most extended sources.

In general we found that a pixel threshold of 30 and an island threshold was most effective in picking up most of the sources while retaining the diffuse emission in the catalogue.

We take cutouts of the sources with the sizes and centre of the cutouts being estimated from the \texttt{PyBDSF} output, and may therefore be imperfect. We manually classify a subset of these sources, labelling them as ``typical'' or ``exotic''. During manual inspection, we flag images that have cutouts that may need to be removed, as they require adjustment or are point sources. An excerpt from the full table we have used is shown in Table~\ref{tab:sourcelist}. 

We do not further adjust these cutouts in our labelled sample in order to keep it consistent with the unlabelled data for training purposes, as well as to be representative of a real-world use case, where large quantities of data would need to be processed in an automated manner. However, adjusted cutouts are made for the released dataset. An exploration of the pybdsf catalogue properties of these sources shows the largest part of the sources are classified as unresolved sources, as expected, giving an average source size of approximately the resolution of the survey, about $8"$ and the maximum source size which we see is almost 12'. We note here that pybdsf is prone to divide large sources in different islands in some cases. Our chosen pybdsf threshold translates to a minimum total flux density of around $50\mu$Jy, meaning that our chosen sources are well above the image noise floor and therefore reliable. The image noise itself varies from image to image but is typically around 2-5 $\mu$Jy/beam. Noise artefacts around bright sources increase the local rms considerably; though we make no specific effort to remove artefacts as we wish to test the robustness of the  VQ-VAE method in the presence of such issues, noting that for an SKA-precursor survey such artefacts would be expected to be fairly common. We remove all the pybdsf-identified unresolved sources from the catalogue. We further remove (using visual inspection) those sources marked as extended but which are essentially single-gaussian sources without much structure from the manually labelled dataset.

Our final sample consists of 13821 unlabelled sources, 2046 labelled typical sources, and 58 labelled exotic sources. We provide the labelled dataset of 2104 sources on Zenodo\footnote{\url{https://zenodo.org/records/20917414}} publicly as a community resource which will be updated with successive data releases adding more classification tags as well as more sources.

We do not perform any augmentation or further pre-processing on the images. This is so that the images may be as close to the true data as possible to check the robustness of the algorithm against data which includes such artefacts which we expect to be present in real surveys and minimises computational cost on large datasets. We normalise each cutout to a range from -0.5 to 0.5 before it is used in any of the machine learning algorithms.

\section{Methodology}

\subsection{Vector Quantised Variational Auto-Encoders}

We use the VQ-VAE implementation created by Sayak Paul\footnote{\url{https://github.com/keras-team/keras-io/blob/master/examples/generative/vq_vae.py}} for Keras\footnote{Chollet, Fran\c{c}ois and others, \url{https://keras.io}} as described in~\citet{vqvae}. The encoder consists of three convolutional layers followed by a vector quantiser, and the decoder similarly consists of three convolutional layers.

Training is done on a training dataset over 50 epochs with a batch size of 256. First, we train the VQ-VAEs on typical sources drawn from the labelled data, under the assumption that the model that was not exposed to exotic sources will be better at reconstructing typical sources compared to exotic sources. A score is subsequently assigned to each image in the test set based on the reconstruction error. We expect a trained VQ-VAE model to produce lower reconstruction errors for typical sources. Thus, we associate poorer reconstruction with more anomalous sources, and use a threshold to select exotic sources.

As a separate experiment, we also train the VQ-VAEs on mixed, unlabelled sources, expected to contain both typical and exotic sources. This allows us to use a much larger total number of sources for training with the aim of improving model performance. As the fraction of exotic sources found in the data through manual inspection is very low (about 2.8 per cent), we expect the normal sources to dominate the learned data distribution. As such, we expect exotic sources to yield higher reconstruction errors due to their natural scarcity, despite training the VQ-VAE on mixed data. The manually labelled data is not used during training, but is used at test time to verify the ability of the fully unsupervised VQ-VAE to discriminate between typical and anomalous sources.

To compare the semi-supervised approach (training on typical sources only) to the unsupervised approach (training on all unlabelled data), we consider three configurations as per Table~\ref{tab:configs}. In configuration A, we train the model with typical sources only, while B and C use unlabelled, mixed data to train in a fully unsupervised manner. We can more directly compare the results from A and B, as both use testing sets in which the number of typical and exotic sources are equal, while C provides a more realistic use case, in which the typical sources far outweigh the exotic sources.

\begin{table}
\begin{tabular}{llll}
\hline
{Sourcename} & {RA (J2000)}   & {Dec (J2000)}  & {Classification} \\ \hline
J1238-4813                       & 12h38m31.2s                     & -48d13m15.6s                     & TYPICAL                             \\
J2250-1632                       & 22h50m13.68s                    & -16d32m20.4s                     & TYPICAL                             \\
J1535-4632                       & 22h50m13.68s                    & -46d32m24s                       & TYPICAL                             \\
J1419-5417                       & 14h19m19.2s                     & -54d17m13.2s                     & EXOTIC                              \\
J0005-2429                       & 00h05m47.52s                    & -24d29m31.2s                     & EXOTIC  \\                 
\hline
\end{tabular}
\caption{An excerpt from the full table of labelled data used in this study, with the first column giving the source name, second and third columns giving the centre of the cutout image in Right Ascension and Declination (J2000 epoch) and the classification between "Exotic" and "Typical" sources given in the final column . The full table of 2104 sources is provided on Zenodo in form of a "live" repository, with separate data releases marking adding more sources and tag specificity. }
\label{tab:sourcelist}
\end{table}

\begin{table}
    \begin{tabular}{p{0.2\linewidth}p{0.2\linewidth}p{0.2\linewidth}p{0.2\linewidth}}
         \hline 
         Configuration & Training Set & Exotic Testing Images & Typical Testing Images\\ \hline
         A & 1991 labelled typical images & 58 & 58 \\
         B & 13824 unlabelled images & 58 & 58 \\
         C & 13824 unlabelled images & 58 & 2046 \\
         \hline
    \end{tabular}
    \caption{The three configurations in which the VQ-VAEs were trained and tested, listing the training data as well as the number of each type of source used in the testing sets.}
    \label{tab:configs}
\end{table}

\subsection{Memory Unit Autoencoders}

Memory Units were introduced in \citet{memae} as a means to prevent autoencoders from generalising anomalous sources that are too similar in structure to typical sources found in the training data. This is accomplished by introducing a memory unit between the encoder and decoder units of the autoencoder. The memory unit consists of several memory slots which contain prototypical patterns that are learned during training. When an input is passed through the model, it is first encoded by the encoder. The code form is then reconstructed from the memory slots. This is then passed to the decoder to reconstruct the original input. As the memory slots are only updated during training, they "memorise" the typical features of the normal data. This reduces performance when reconstructing anomalous data and increases the reconstruction error for such sources. 

Memory unit AEs where applied to astronomical data by \citet{Brand2026} which applied numerous models to anomaly detection in FIRST data. Among the models that were compared, memory AEs where the best at anomaly detection, excluding combinations of models. We take their implementation of a memory unit AE\footnote{\url{https://github.com/KBrand26/CARA}}, which we will call the CARA Memory AE, and compare it to the VQ-VAE on our dataset. We run CARA Memory in the same manner as configuration B.

\subsection{Astronomaly Protege}

\texttt{Astronomaly}\footnote{\url{https://github.com/MichelleLochner/astronomaly}} is a anomaly finding software, described in \citet{Lochner2021}. \texttt{Astronomaly} utilises a "human in the loop" approach in which the system will score provided sources according to feedback provided by the researcher. \texttt{Astronomaly} assigns each cutout a score from 0 to 5, with 5 being the most anomalous, creating a ranking of sources. The user is presented with the highest scoring source and can then provide their own score. \texttt{Astronomaly} will learn from these user-provided labels and can then be retrained to provide a new ranking. 

We compare an extension of \texttt{Astronomaly} called \texttt{Protege}\footnote{\url{https://github.com/MichelleLochner/mgcls.protege}}, which has been applied to MGCLS data \citep{Protege} and which makes use of Bootstrap Your Own Latent \citep[BYOL,][]{byol} for feature extraction. After the initial feature extraction, \texttt{Astronomaly} allows the user to score various sources and then retrains to provide an updated ranking. BYOL is only used for the inital feature extraction after which \texttt{Protege} will attempt to predict user score based on these features. Due to the difference in how \texttt{Astronomaly Protege} works it is difficult to make a direct comparison. We do the initial training of the feature extractor using the unlabelled data. Testing is done with all labelled sources. Some fraction of these labels are provided to \texttt{Astronomaly} which then retrains and reorders the sources based on the user score. We provide a user score of 5 for sources labelled exotic and a user score of 1 for sources labelled typical.

We provide \texttt{Astronomaly} woth 1, 5 and 10~percent of labelled data and plot ROC curves for each by varying the threshold score at which we call a source exotic.

\subsection{Performance Evaluation}

Sources correctly identified as exotic are referred to as true positives (TP). Similarly, those correctly identified as typical sources are referred to as true negatives (TN), and those incorrectly identified as exotic or typical false positives (FP) and false negatives (FN), respectively.

\begin{figure}
\centering
\includegraphics[width=0.5\textwidth]{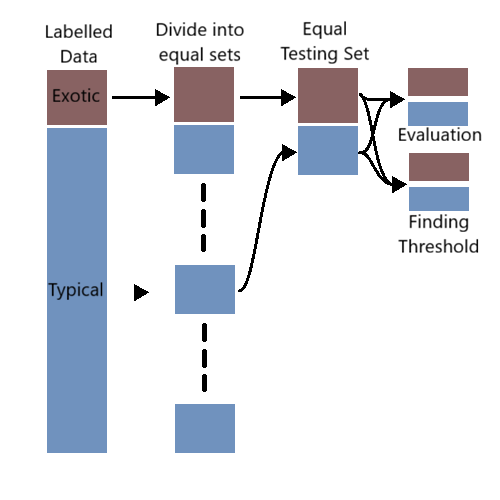}
\caption[Data Structure for Evaluation]{Structure of the data for evaluation of models in configurations A and B. Typical sources are split into chunks equal in number to the exotic sources. Exotic labelled sources cannot be split due to their low number. For each run, a different chunk of typical sources is used for testing so that each is used once. In configuration A the remaining typical sources are used for training. The testing set is then split into a set used to find the optimal threshold and another set to evaluate the model at that threshold.}
\label{fig:data}
\end{figure}

The models are trained and tested 35 times each to provide a better statistical understanding of their performance. For configurations A and B, we test the models with equal numbers of typical and exotic sources. As the number of exotic sources is very limited, even in a survey as large as MGCLS, we use the entire set of 58 labelled exotic sources each time. The labelled typical sources are split into sets of equal size, each set having 58 sources to match the number of exotic sources, resulting in 35 sets. We use each typical subset in testing exactly once. In configuration A, this also means that each subset is excluded from the training set exactly once. This process is illustrated in Figure~\ref{fig:data}. The model is retrained and retested for each of these sets. For configuration A, the remainder of the labelled typical sources are used to train the model for each of the 35 sets.

In order to evaluate how well the VQ-VAE has reconstructed a particular image, the input and output images are compared by calculating the normalised cross-correlation (NCC). NCC is a widely used similarity measure that is similar to convolution, and is somewhat invariant to changes in brightness and contrast. We select NCC to calculate the score as it will be less sensitive to differences in intensities and background variation, as well as being lenient to translational differences. This allows for scoring morphology more directly while being less sensitive to other factors. This leniency is best suited to images that may not be centred or have their backgrounds and artefacts removed. NCC produces a score between 0 (worst reconstructed) and 1 (perfectly reconstructed). We treat sources with lower scores as more anomalous. For two 2-dimensional arrays, $I$ and $K$, of $m$ by $n$ pixels, it is given by

\begin{equation}
(I*K)(i,j) = \sum_{m}^{}\sum_{n}^{}I(i+m,j+n)K(m,n)\label{eq:cc}.
\end{equation}

NCC is used in order to account for variations in intensity and background noise that should ideally not affect the outcome of the similarity measurement. The NCC measure for two images $f$ and $t$ with dimensions $m\times n$ is 

\begin{equation}
NCC = \frac{1}{mn}\sum_{x}^{}\sum_{y}^{} \frac{1}{\sigma_{f}\sigma_{t}} (f(x,y) - \mu_{f})(t(x,y) - \mu_{t})\label{eq:ncc},
\end{equation}

\noindent where \(\sigma\) and \(\mu\) are the standard deviation and mean of each image, respectively. The value is averaged over all pixels, and is equal to $1$ only if the images are identical. We use the NCC between the original and reconstructed images as our similarity score, with more exotic sources being expected to have a lower NCC score.

We can use the testing set to create a receiver operating characteristic (ROC) curve for a particular model. ROCs are a widely used metric in binary classification, and constitute a plot between the false positive rate (FPR) and true positive rate (TPR) at varied binary classification threshold values. This is the threshold value at which a reconstruction score indicates a source to be exotic rather than typical. TPR, also called recall, is the fraction of positive results that are true TPs. FPR is the fraction of negative samples that are reported as FPs. In case of perfect classification, FPR should be 0, while TPR should be 1. The area under this curve (AUC) is a common metric of success for binary classifiers. 

Another common performance metric for a binary classifier is the harmonic mean of the precision and recall, or F-Score. The \(F_{1}\) score represents the harmonic mean of these two quantities in which they are weighted equally, although a more general form, \(F_{\beta}\), applies different weights to the two metrics. In our particular application, it may be more beneficial to favour recall over precision, and contend with more false positives in order to locate more scientifically interesting unusual sources. Therefore, we test both the more common \(F_{1}\) score, as well as the \(F_{2}\) score, in which recall is given twice the weight of precision.

In order to determine the appropriate anomaly classification thresholds, we split the testing into two smaller equal sets. The first is used to determine the best thresholds for \(F_{1}\) and \(F_{2}\). The second is used to determine the \(F_{1}\) and \(F_{2}\) scores given those thresholds and construct a confusion matrix.

To better assess the ability of VQ-VAE to separate the anomalous from normal sources in its internal (latent) representation, we visualise the resulting latent spaces using uniform manifold approximation and projection\footnote{\url{https://github.com/lmcinnes/umap}} \citep[UMAP,][]{umap}. This technique is designed to visualise high-dimensional data by projecting it down into a 2-dimensional representation. UMAP models the high level data as a manifold and projects this manifold down to a lower dimensional space while preserving topological structures. It does this by constructing a weighted graph for the data by assigning probabilistic edge weights to nearest neighbours in order to create a probabilistic fuzzy graph. It then creates a random initial lower dimensional representation and optimises it by using gradient descent to minimise the cross-entropy between the high and low dimensional graphs. This forces those points that are close in the higher dimensional graph to cluster on the lower dimensional projection. By projecting the encoded latent space representations of the testing images onto a 2D plane, we may examine the separation of the encoded forms of the typical and exotic sources.

\section{Results}

\subsection{VQ-VAE Performance in Three Configurations}

We test the VQ-VAE in the three configurations summarised in Table~\ref{tab:configs}.

The test set for configuration A is composed of an equal number of exotic and typical sources drawn from the labelled data. The remaining typical sources from the labelled data form the training set. Due to the low number of labelled exotic sources, all of the exotic data are used to test the model during each run. An equal number of typical sources are drawn from the labelled data. A different subset of typical sources are used for each run. The model is trained and tested 35 independent times, using a different set of the labelled testing data each run so that each labelled typical source is in a testing set once. As such, the VQ-VAE is trained on labelled, typical sources only.

\begin{figure}
\centering
\includegraphics[width=0.49\textwidth]{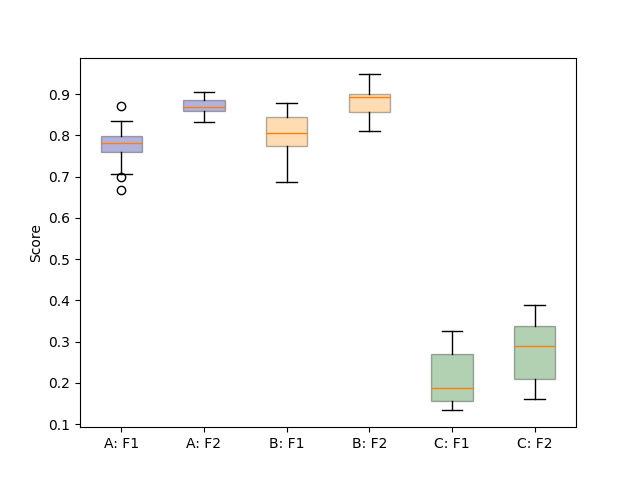}
\caption[$F_1$ and $F_2$ Scores in configurations A, B and C]{Comparison of the distribution of $F_1$ and $F_2$ scores for models in configurations A, B and C as per Table~\ref{tab:configs}. Models in configuration A are trained on labelled data while those in configuration B are trained on the unlabelled data. Both are tested with testing sets having an equal number of exotic and typical sources. Models in configuration C are trained on unlabelled data and tested with an asymmetric testing set. The five number summary for each distribution shows the median in the centre of the box, with half of all values falling within the box. The whiskers show the minimum and maximum, excluding the outliers which are marked by circles.}
\label{fig:f1f2comp}
\end{figure}

For this configuration with equally large typical and exotic testing sets, the expected $F_1$ and $F_2$ scores of a random guess would be $0.5$. The AUC of a random guess would also be $0.5$. Figure~\ref{fig:f1f2comp} shows the distribution of the actual $F_1$ and $F_2$ scores from testing. The mean $F_1$ score was $0.7762$ with a standard deviation of $0.0407$ and a median of $0.7826$. The mean $F_2$ score was $0.8720$ with a standard deviation of $0.0178$ and a median of $0.8696$.

\begin{figure}
\centering
\includegraphics[width=0.49\textwidth]{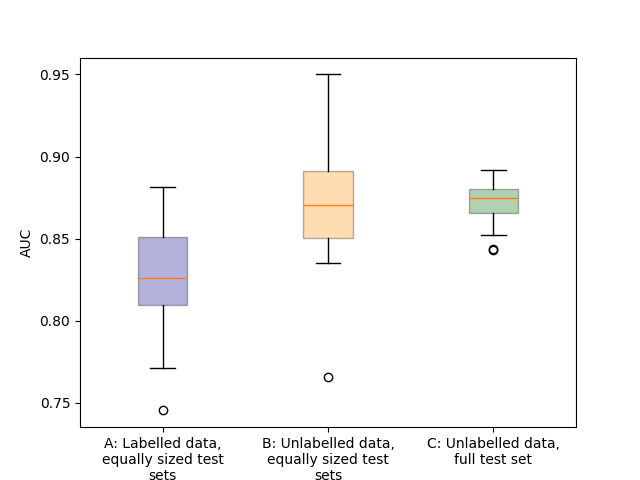}
\caption[AUC of VQ-VAE in all configurations]{(Left) Box and whisker diagram for AUC in configuration A (Centre) Box and whisker diagram for AUC in configuration B (Right) Box and whisker diagram for AUC in configuration C. The five number summary for each distribution shows the median in the centre of the box, with half of all values falling within the box. The whiskers show the minimum and maximum, excluding the outliers which are marked by circles.}
\label{fig:aucbox}
\end{figure}

\begin{figure*}
\centering
\includegraphics[width=0.33\textwidth]{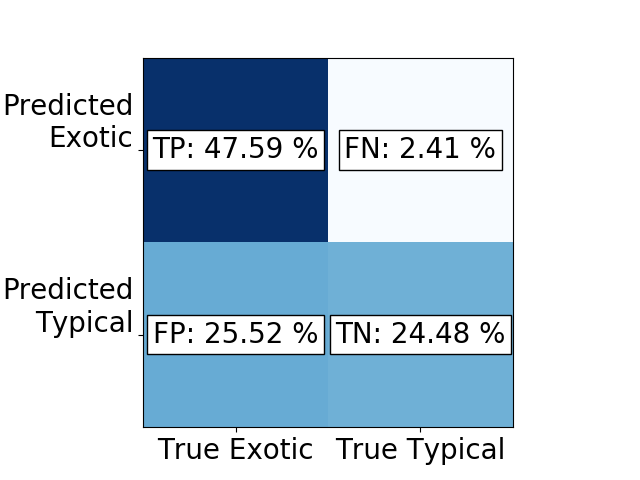}
\includegraphics[width=0.33\textwidth]{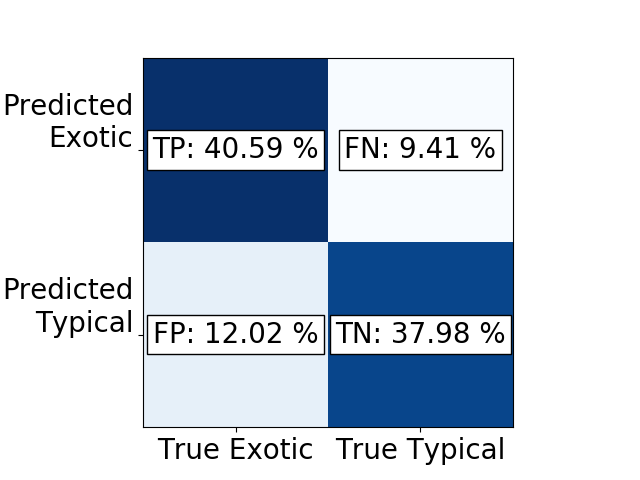}
\includegraphics[width=0.33\textwidth]{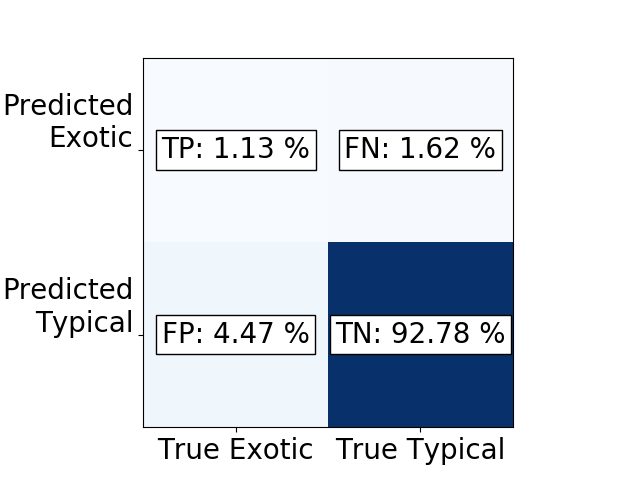}
\caption[Confusion Matrices of the various configurations]{Averaged confusion matrix of the model in configuration A, trained on labelled typical images with an equal number of typical and exotic images in the test set (left). Averaged confusion matrix of the model in configuration B, trained on all unlabelled images with an equal number of typical and exotic images in the test set (centre). Averaged confusion matrix of the model in configuration C trained on all unlabelled images with all labelled images being used for the test set (right).}
\label{fig:confm}
\end{figure*}

As shown in Figure~\ref{fig:aucbox}, the mean AUC for the ROC was $0.8270$  with a standard deviation of $0.0298$. As such, we conclude that the model manages to discriminate between the types of sources. The averaged confusion matrices are shown in Figure~\ref{fig:confm}.

\begin{figure}
\centering
\includegraphics[width=0.49\textwidth]{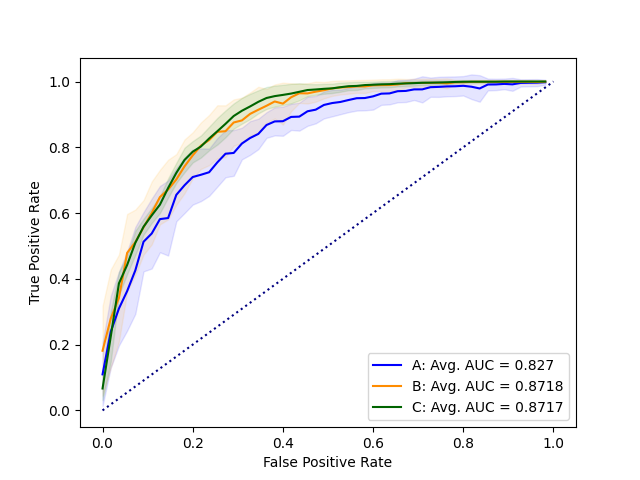}
\caption[Average ROC of the VQ-VAE in three configurations]{Average ROC with AUC of the VQ-VAE in configurations A, B and C. The shaded region represents the standard deviation in the curve.}
\label{fig:rocvqvae}
\end{figure}

Figure~\ref{fig:rocvqvae} shows the average ROC with standard deviation of these tests. We can see that the ROC sits well above the line that a random class selection would follow, as suggested by the AUC. In line with the F-scores and the AUC, the ROC suggests that the model is indeed able to distinguish between the sources.

\begin{figure*}
\centering
\includegraphics[width=0.34\textwidth]{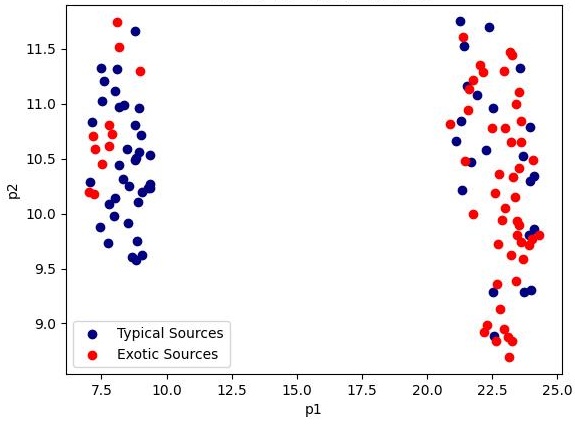}
\includegraphics[width=0.33\textwidth]{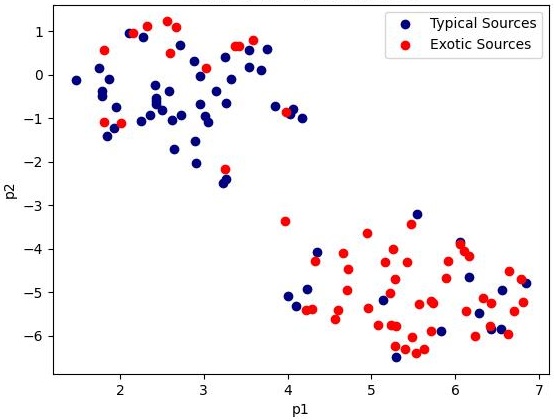}
\includegraphics[width=0.32\textwidth]{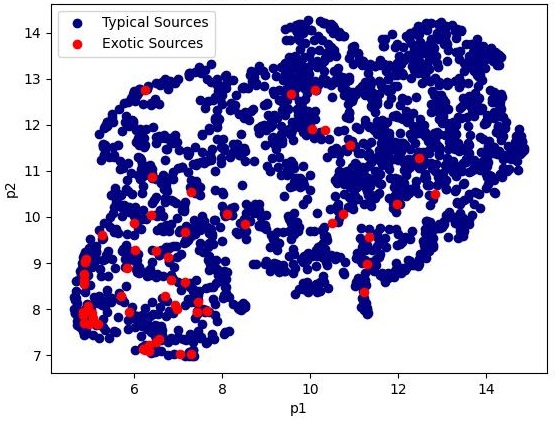}
\caption[Selected UMAPs VQ-VAE Latent Spaces]{(Left) Configuration A: UMAP of the code forms of the testing set from a selected run of the VQ-VAE trained with labelled typical sources. The test set consists of an equal number of typical and exotic sources, with the UMAP mappings of their code forms show in blue and red dots respectively. (Centre) Configuration B: UMAP of the code forms of the testing set from a selected run of the VQ-VAE trained with unlabelled sources. The test set consists of an equal number of typical and exotic sources, shown in blue and red dots respectively. (Right) Configuration C: UMAP of a selected run of the VQ-VAE in configuration C trained with unlabelled data and tested on the full labelled dataset. The mappings of the code forms of the typical and exotic labelled data are shown in blue and red, respectively.}
\label{fig:umap}
\end{figure*}

We further analyse the efficacy of VQ-VAE by selecting one of the runs, and inspecting its latent space using UMAP 2D projection. Figure \ref{fig:umap} shows the UMAP mappings of the code forms of the test set. UMAP shows two distinct clusters, corresponding to the typical and anomalous sources. While there is some overlap between the clusters, it is clear that some degree of separation has been obtained. It is evident from Figure~\ref{fig:umap} that VQ-VAE is more likely to misclassify a typical source as an exotic (FP) than an exotic as a typical (FN), which is a desirable property.

As such, although VQ-VAE shows initial promise, the performance of the model is not sufficient to be deployed in a fully autonomous fashion without manual inspection. Indeed, that is the intention: that expert astronomers take over to assess candidate cases in a fraction of the total number of sources.

In configuration B, we train the model on all of the unlabelled data. The model is then tested on an equal number of typical and exotic sources from the labelled data, in the same manner as configuration A.

Figure~\ref{fig:f1f2comp} shows how the $F_1$ and $F_2$ scores differ from those of the models trained on labelled data. The mean $F_1$ score is $0.8044$ with a median of $0.7857$ and a standard deviation of $0.0462$. The mean $F_2$ score is $0.8795$ with a median of $0.9032$ and a standard deviation of $0.0363$. We see that the performance has improved slightly compared to configuration A. This is likely due to using a significantly larger training set. Figure~\ref{fig:aucbox} provides a comparison of the AUC for the different configurations, once again confirming a noticeable improvement in performance. The mean AUC for configuration B is $0.8711$ with a median of $0.8701$ and a standard deviation of $0.0291$. 

Although the training data contains approximately $2.8\%$ exotic sources (based on the ratio observed in the labelled set), a significantly increased total number of sources to train on leads to a significant improvement in source reconstruction. Since the vast majority of sources are typical, reconstruction of the atypical sources remains a harder task. Figure~\ref{fig:rocvqvae} shows the average ROC for configuration B. As expected from the AUC, we see that the ROC indicates good performance well above random, with slighly better AUC that that of configuration A. Figure~\ref{fig:umap} show the testing data projected onto the latent space for one run of the configuration using UMAP. Similarly to configuration A, we can see some clustering of the typical and exotic sources by type.

We also perform the Mann–Whitney U \citep{mwu} test on the ROC values to compare the increase in performance from training on the larger data set of unlabelled sources over that of training on the smaller set of labelled sources. The null hypothesis is that the medians of the sets of $F_1$ scores produced by these two methods are the same. 
The p-value of the test is $0.00034$, indicating that the result is significant at p < 0.05. As such, there is a statistically significant improvement in performance when training on a larger set of unlabelled data.

The model in configuration B significantly improves upon the performance of configuration A. While the results may still require a post-hoc manual inspection, the burden of labelling a subset of sources as typical prior to training is alleviated in this setup. Configuration B clearly demonstrates that the larger training set more than makes up for the presence of exotic sources in VQ-VAE training. Training on unlabelled, larger datasets seems to be a promising direction.

For the final configuration, C, we train the model on all unlabelled data, and the testing set consists of all labelled data. The model itself is identical to the second configuration,  but is tested in a more realistic manner, with a testing set that more closely resembles the distribution of sources found in real data. This means that the training and testing sets are the same for each run.

We run this configuration 35 times as well to account for stochasticity arising from random model initialisation and data subsampling during training (i.e., mini-batching). All of the data is used in each run, but shuffled so that the same data is not used for finding thresholds and calculating test metrics.

The labelled data consists of $2.8\%$ exotic sources, which gives the expected $F_1$ and $F_2$ scores of a random guess of $0.052$ and $0.028$, respectively.

Figure~\ref{fig:f1f2comp} shows the distribution of $F_1$ and $F_2$ scores for configuration C. The mean $F_1$ score is $0.2107$ with a median of $0.1880$ and a standard deviation of $0.0629$, significantly better than random class selection. Likewise, the mean $F_2$ score is $0.2714$ with a median of $0.2900$ and a standard deviation of $0.0709$. These scores are significantly better than random, but are still low enough to show how difficult it can be to extract such a small fraction of exotic sources. This configuration puts the results from configuration B into context, and demonstrates the complexity of the real-world scenario. The models may be a useful tool, but cannot be used in isolation or without expert intervention.

Figure~\ref{fig:aucbox} shows the AUC compared to that of configurations A and B. We see that the performance indicated by the AUC is very similar to that of configurations A and B, although with a smaller deviation. This is most likely due to the difference in testing sets. Similarity between AUCs of configurations B and C is expected, as they are trained on the same data. 

The average ROC of these runs is shown in Figure~\ref{fig:rocvqvae}, where the mean AUC is $0.8717$ with a standard deviation of $0.0118$. As expected, it is quite similar to that of  configuration B.

The UMAP visualisation for a single run of configuration C is shown in Figure~\ref{fig:umap}. Although some clustering of the exotic and typical sources is evident, the separation is not clear or linear. As in Configuration A, we can see that even a small fraction of the typical sources can create a large overlap with the mappings of the anomalous sources. This can represent a large number of false positives, necessitating human inspection.

\subsection{Comparison with Simple AE}

\begin{figure}
\centering
\includegraphics[width=0.49\textwidth]{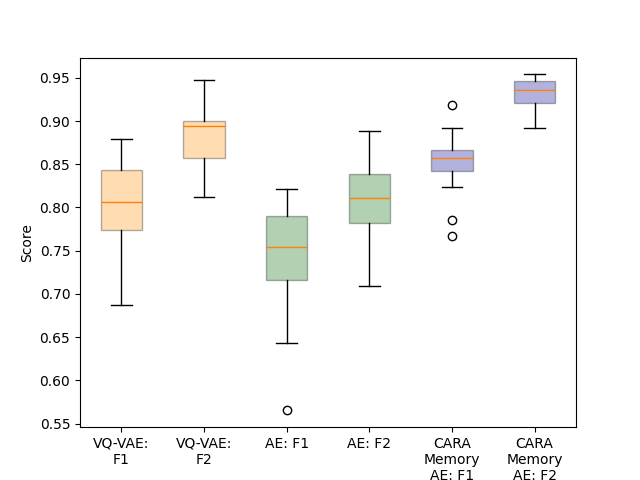}
\caption[Distribution of $F_1$ and $F_2$ scores for the three AEs]{Distribution of the $F_1$ and $F_2$ scores for the three autoencoders over 35 runs in configuration B. On the left is the VQ-VAE with mean $F_1$ score of 0.8044, and a mean $F_2$ score of 0.8795. In the centre the simple AE with a mean $F_1$ being 0.7482 and mean $F_2$ being 0.81497. On the right in blue, CARA Memory AE with a mean $F_1$ of 0.8504 and a mean $F_2$ of  0.9307}
\label{fig:aefscores}
\end{figure}

\begin{figure}
\centering
\includegraphics[width=0.49\textwidth]{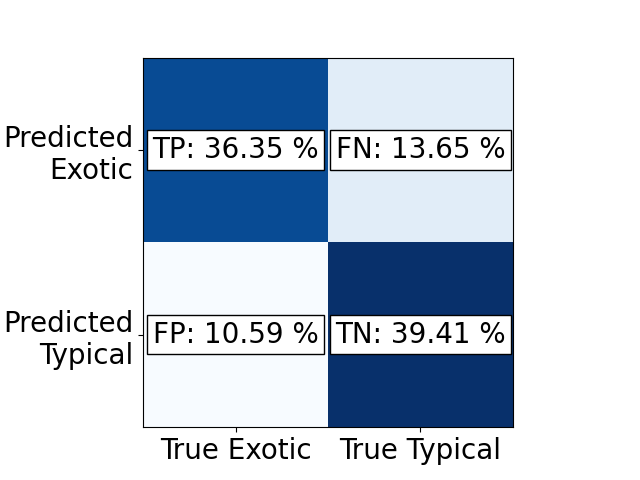}
\caption[AE Confusion Matrix]{Confusion matrix for a simple AE over 35 runs in configuration B with mean $F_1$ being 0.7482 and mean $F_2$ being 0.81497.}
\label{fig:aeconfm}
\end{figure}

We compare the performance of the VQ-VAE to a standard AE. For a direct comparison, our AE model simply consists of the same model with the vector quantiser layers turned off. We compare the two models in configuration B. The distribution of $F_1$ and $F_2$ scores is show in Figure~\ref{fig:aefscores} and we note the average $F_1$ of 0.7482 as opposed to 0.8044 for the VQ-VAE. Similarly the average $F_2$ score is 0.81497 as opposed to 0.8795 for the VQ-VAE. The averaged confusion matrix for the AE is given in Figure~\ref{fig:aeconfm}.

\begin{figure}
\centering
\includegraphics[width=0.55\textwidth]{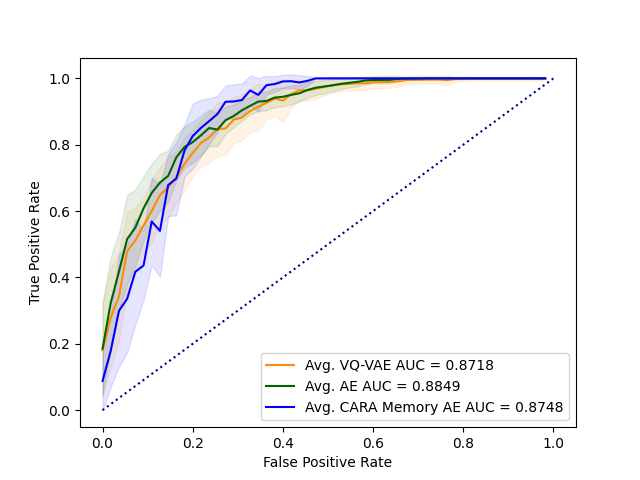}
\caption[Average ROCs of the three compared models]{Average ROCs with shaded standard deviation of the AE, VQ-VAE and CARA Memory AE models trained in configuration B with standard deviation shaded.}
\label{fig:aerocs}
\end{figure}

We show the average ROC of the AE overlaid with the average ROC of the VQ-VAE in configuration B in Figure~\ref{fig:aerocs}. Despite the lower $F$-scores, the ROCs are very comparable with similar AUCs. The higher $F_1$ and $F_2$ scores show that the VQ-VAE maintains an advantage over the simple AE for this task.

\subsection{Comparison with CARA Memory AE}

\begin{figure}
\centering
\includegraphics[width=0.49\textwidth]{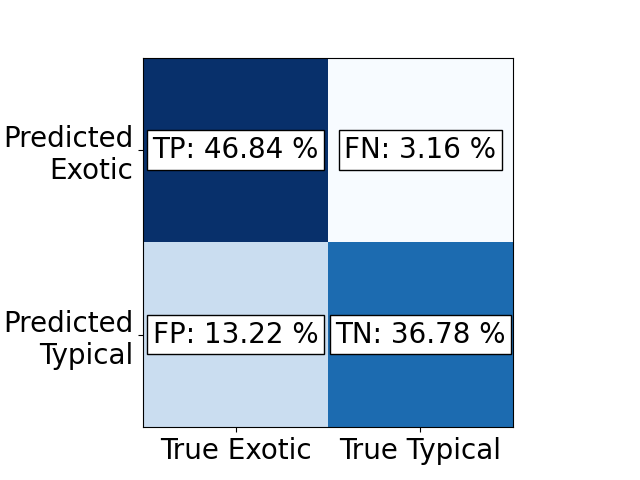}
\caption[CARA Memory AE Confusion Matrix]{Confusion matrix for the CARA Memory AE over 35 runs in configuration B with mean $F_1$ being 0.8504 and mean $F_2$ being 0.9307.}
\label{fig:caraconfm}
\end{figure}

We compare the performance of the CARA Memory AE model in configuration B. Figure~\ref{fig:aefscores} shows the distribution of $F$-scores with a mean $F_1$ of 0.8504 and a mean $F_2$ of  0.9307 and the averaged confusion matrix is in Figure~\ref{fig:caraconfm}. These values show a meaningful improvement over the VQ-VAE model. The confusion matrix seems to suggest that most of the performance gain comes from a much reduced false negative rate, even with a slightly higher false positive rate.

The ROC is given in Figure~\ref{fig:aerocs}. Despite the improvement in the $F$-scores the AUC is very similar to those of the VQ-VAE and simple AE. Noticably, it behaves a little differently in having a slightly sharper increase in TPR as the FPR increases. However, it has a very similar AUC.
The improved performance of the Memory AE does come with a significant performance penalty, however, taking several times longer to train on the same hardware. This is a significant trade-off for larger datasets.

\subsection{Comparison with Astronomaly}

Finally, we examine the performance of \texttt{Astronomaly} on the same data. \texttt{Astronomaly} cannot be directly compared as it uses a human-in-the-loop approach that attempts to learn the preferences of the operator in identifying anomalous sources. For this reason it is presented only as a subjective comparison to provide context of a commonly used astronomy tool for finding anomalies.

\begin{figure}
\centering
\includegraphics[width=0.49\textwidth]{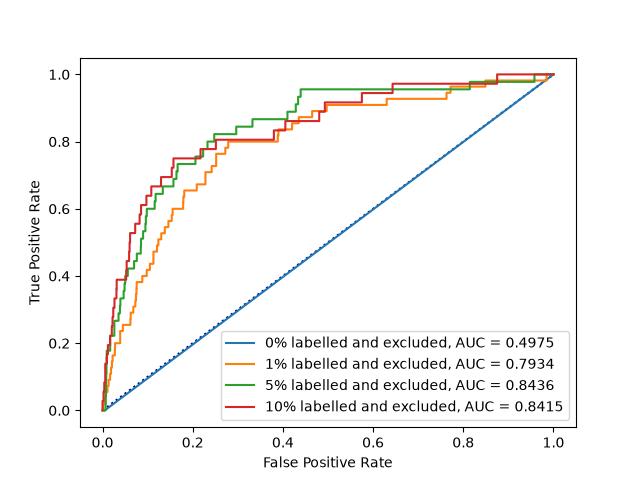}
\caption[Astronomaly ROC curves]{Astronomaly ROC curves with given fractions of sources labelled}
\label{fig:astronomaly}
\end{figure}

We train the BYOL feature extractor on the unlabelled data and then test it on our labelled dataset. The entire labelled dataset is used for testing. 
We test \texttt{Astronomaly} with 1, 5 and 10~percent of labels being provided to the software. \texttt{Astronomaly} accepts an anomaly label in the range of 0 to 5. We label sources either 1 or 5 depending on our labelled dataset. After new sources are labelled, \texttt{Astronomaly} retrains then re-orders the sources based on the new labels. BYOL also takes significantly longer than the VQ-VAE to run but reordering the sources is fast. We generate an ROC for the ordered sources by varying the index in the list that is treated as a threshold. These ROC curves are shown in Figure~\ref{fig:astronomaly} with their AUC values. The maximum AUC is slightly lower but comparable to those of the VQ-VAE but requires a significant fraction of the sources to be labelled.

\subsection{A Sample of Automatically Selected Anomalous Sources}

To illustrate the types of sources that may be found in the MGCLS data using VQ-VAE, we select some of the sources from the exotic, typical and unlabelled datasets and show them along with their reconstructions. We manually select five samples from the sources with the highest NCC (a score that indicates that the model reconstructed the source well, thus the source is more typical) and those with the lowest NCC (a score indicating that the model reconstructed the source poorly, suggesting that the source is anomalous) for each type of source. The NCC scores were determined by a particular run of a model in configuration C. 

\begin{figure*}
\centering\includegraphics[width=1.05\textwidth]{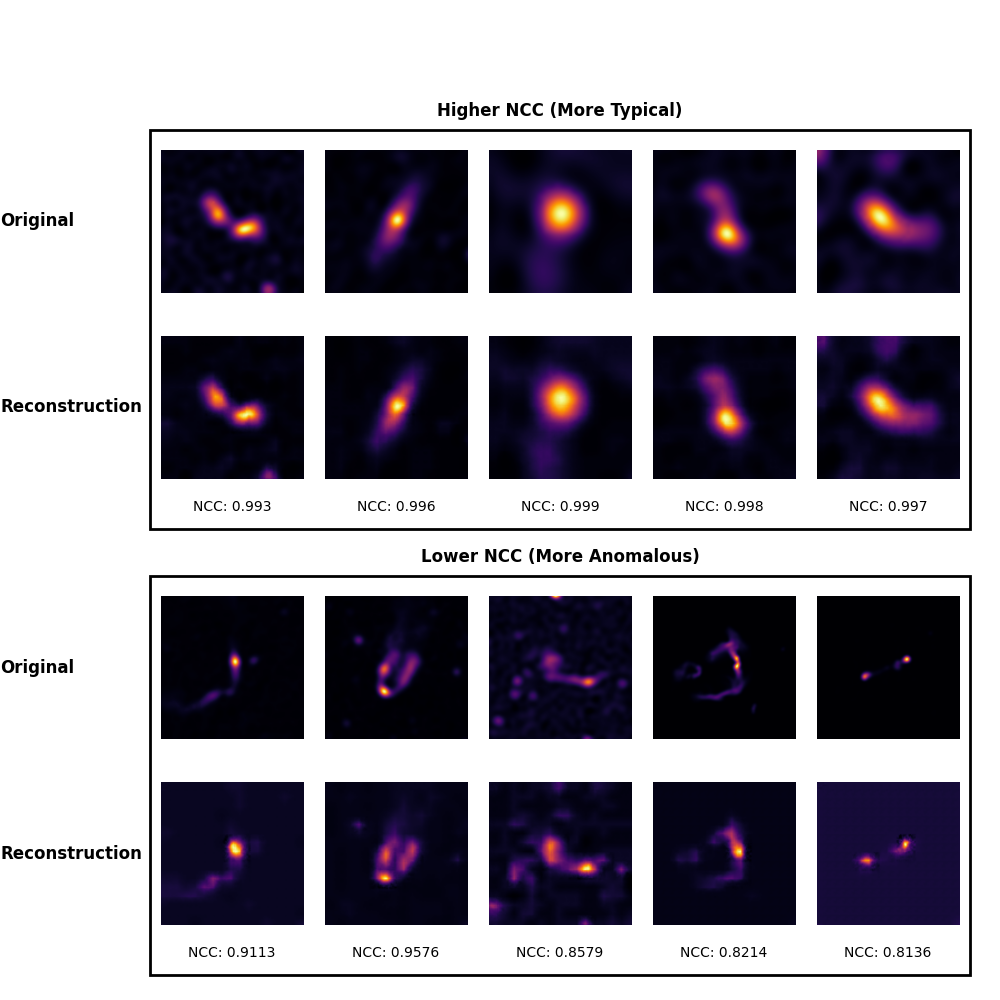}
\caption[A sample of the best and worst reconstructed typical sources.]{A sample of the best and worst reconstructed typical sources with their reconstructions as performed by a VQ-VAE trained on the unlabelled data. (Top) Five of the best reconstructed typical sources, having the highest scores and therefore ranked as "most typical." They are shown with their reconstructions as well as the reconstruction score. (Below) Five of the worst reconstructed typical sources, having the lowest scores and therefore ranked as "least typical". These sources come up as false positives. They are shown with their reconstructions and reconstruction scores.}
\label{fig:typ_recon}
\end{figure*}

In Figure~\ref{fig:typ_recon}, which shows sources manually labelled as typical, we can see true negatives with higher NCCs, as well as some false negatives which have low NCCs. Although all sources in Figure~\ref{fig:typ_recon} were labelled as typical by the astronomers, it is clear that the sources falsely labelled as exotic by VQ-VAE indeed have more nuanced structure. It is important to note that the notion of ``exotic'' is subjective in itself, and what may be ``typical'' to an astronomer may still pose a significant challenge to a model learning source reconstruction. 

\begin{figure*}
\centering
\includegraphics[width=1.05\textwidth]{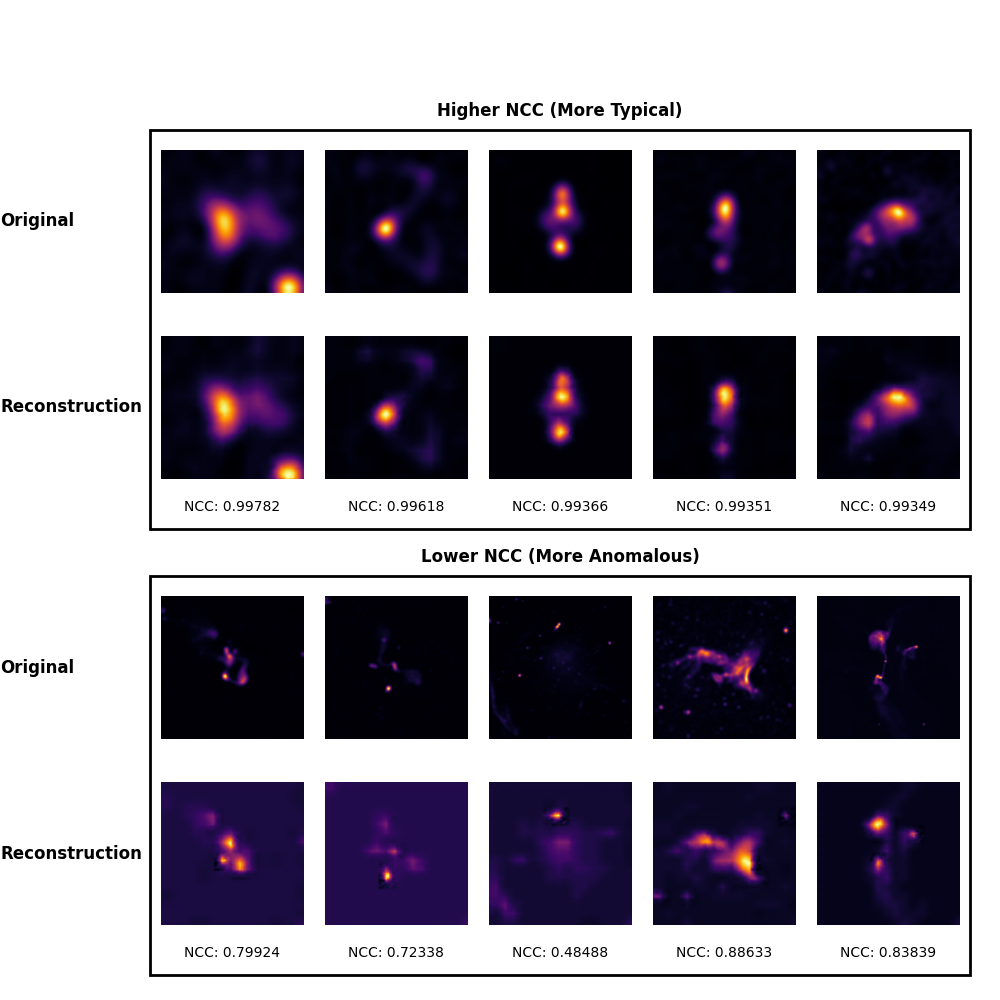}
\caption[A sample of the best and worst reconstructed exotic sources.]{A sample of the best and worst reconstructed exotic sources with their reconstructions as performed by a VQ-VAE trained on the unlabelled data. (Top) Five of the best reconstructed exotic sources, having the highest scores and therefore ranked as "most typical." These sources are false negatives. They are shown with their reconstructions as well as the reconstruction score. (Below) Five of the worst reconstructed exotic sources, having the lowest scores and therefore ranked as "least typical". They are shown with their reconstructions and reconstruction scores.}
\label{fig:exo_recon}
\end{figure*}

In Figure~\ref{fig:exo_recon}, we show sources manually labelled as anomalous. Here, the sources with the lower NCCs represent true positives, while those with higher NCCs are false negatives, and would be lost amongst the typical sources if classified automatically using VQ-VAE. Notably, the false negatives can all be described as ``fuzzy'' and blob-like. Supposedly, it is the lack of high-definition intricate structure around the sources that prevented the model from identifying them as exotic.

\begin{figure*}
\centering
\includegraphics[width=1.05\textwidth]{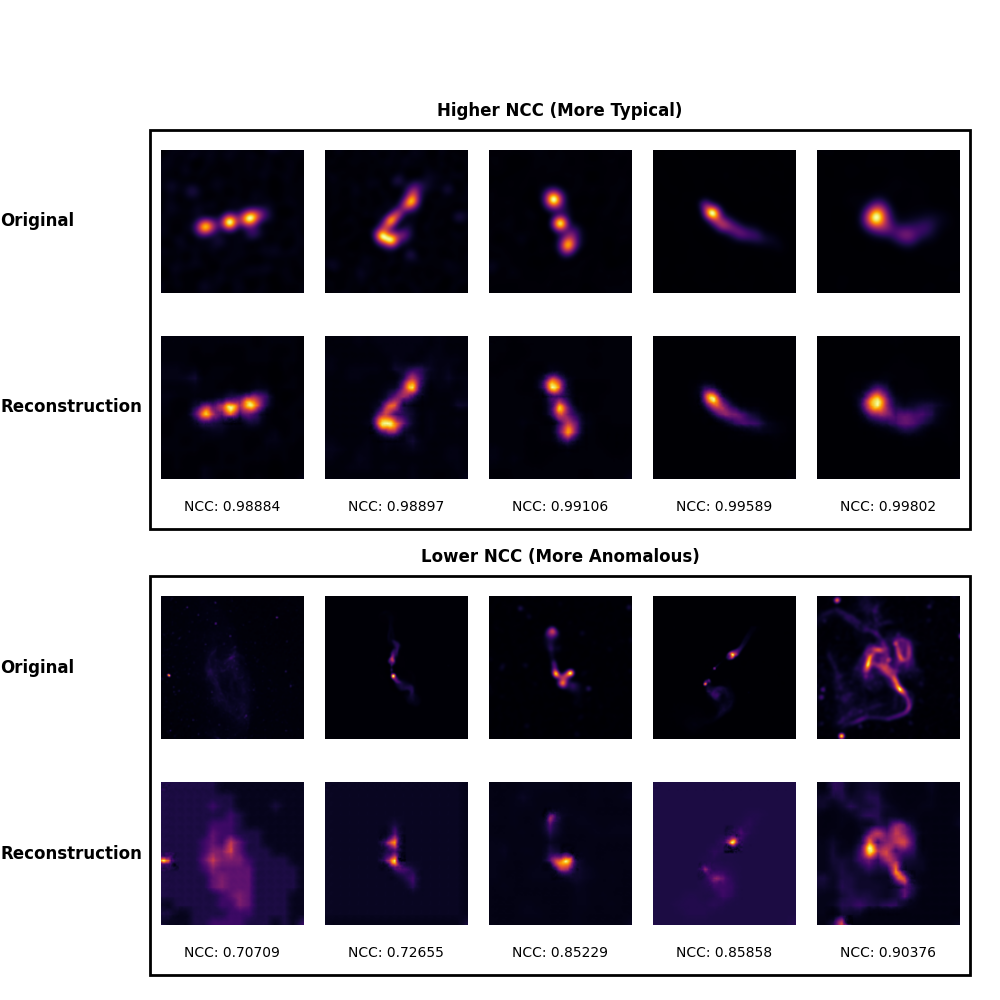}

\caption[A sample of the best and worst reconstructed unlabelled sources.]{A sample of the best and worst reconstructed sources from the unlabelled data with their reconstructions as performed by a VQ-VAE trained on the unlabelled data. (Top) Five of the best reconstructed sources, having the highest scores and therefore ranked as "most typical." They are shown with their reconstructions as well as the reconstruction score. (Below) Five of the worst reconstructed sources, having the lowest scores and therefore ranked as "least typical". They are shown with their reconstructions and reconstruction scores.}
\label{fig:unl_recon}
\end{figure*}

Finally, in Figure~\ref{fig:unl_recon} we show a sample of high NCC and low NCC sources from the unlabelled data. Once again, sources that resemble Gaussian blobs are reconstructed well and are labelled typical, while sources with interesting diffuse emissions were labelled as anomalous. This demonstrates the ability of the model to extract interesting sources from the data, as the more unusual sources indeed tend to have lower NCC scores.

The sources shown in Figures~\ref{fig:typ_recon} to Figure~\ref{fig:unl_recon} also demonstrate the high quality of the MeerKAT data. The sources are good candidates for further study, including looking at multiwavelength data, which is beyond the scope of this paper. Once the most interesting sources have been identified, multiwavelength data for them may be obtained, and various properties as well as their environment may be analysed to find possible reasons for their unusual morphologies.

\section{Conclusion and Outlook}
In this work, we demonstrated the application of vector quantised variational Autoencoders (VQ-VAEs) to 1.28 GHz radio continuum images from the MeerKAT Galaxy Cluster Legacy Survey (MGCLS). By learning a compact, discrete representation of the data distribution, VQ-VAEs can effectively highlight sources that deviate from typical morphologies, making them well-suited for anomaly detection in large astronomical datasets. Our results showed that VQ-VAEs, even when trained in an unsupervised manner, can filter out the majority of common sources while isolating potentially interesting or unusual structures. We also release a labelled test set of MGCLS radio galaxies to support further research.

The performance of the model using a large unlabelled dataset as opposed to a smaller labelled dataset shows that unsupervised learning is not only feasible but preferable when it allows for the use of a much larger training set than labelled data.

In comparing the performance of VQ-VAEs to some other published models, we find that there are trade-offs that may make certain models more applicable to certain situations. VQ-VAEs represent an improvement over simple AEs without too much of a performance penalty. The CARA Memory AE shows better performance than the VQ-VAE in anomaly detection, but with a very significantly higher computation cost that may be unsuitable for large datasets where low computational cost is desired. Looking at the AUCs may suggest that VQ-VAE may perform comparably to \texttt{Astronomaly Protege} in anomaly detection, but is much faster to train than BYOL and doesn't require labelled sources.

The classifications produced by unsupervised VQ-VAE tested in this paper, while not sufficient to be relied upon without further inspection, can significantly reduce the workload of an astronomer faced with tens or hundreds of thousands of sources. For well-selected anomaly thresholds, VQ-VAE returned the most exotic sources and excluded the most typical sources. A fairly high number of false positives resulted from certain sources labelled by astronomers as exotic having blob-like shapes and the lack of finely structured detail. The low number of exotic sources in the data implies that the true positives may still be outnumbered by false positives. The use case and value in VQ-VAE is to help the user automatically search through large collections of potential sources by reducing the total number without discarding the majority of exotic sources. By carefully configuring the anomaly threshold, VQ-VAE may be able to remove a large proportion of the typical sources while retaining almost all of the exotic sources.

Many sources may be classified as exotic by experts because of a deeper understanding of the source itself, while not having features that are particularly unusual in shape. For automated classification, a method that considers other properties of a source, such as multi-wavelength data, may likewise outperform a model that classifies purely in the radio image plane. It should also be noted that the categories used of ``typical'' and ``exotic'' are very broad and inherently subjective. Searching for a more narrowly defined type of morphology, such as X-shaped sources or bent tail sources, may yield stronger results. VQ-VAEs are a promising candidate for further development in this area. We intend to explore these options, as well as apply VQ-VAEs to other large radio surveys, such as the recent MeerKAT survey on the South Pole Telescope (SPT) survey field. As the VQ-VAE is variational, testing a probabilistic score was initially considered and may also be a valuable avenue for future work.

Automated approaches such as those explored in this paper remain key to searching through high volumes of data and extracting the full scientific potential of new and upcoming telescopes such as the SKA and its pathfinders that have massively increased depth, area, imaging fidelity, and dynamic range. In addition, the set of 2104 labelled MeerKAT source images, which we make freely available, forms a resource of considerable value for the community, which we aim to maintain in a "live" form with frequent updates for additional sources and classification tags. 

\section{Acknowledgements}

F.L. Ventura acknowledges financial support from the Inter-University Institute for Data Intensive Astronomy (IDIA). IDIA is a partnership of the University of Cape Town, the University of Pretoria and the University of the Western Cape. IDIA is registered on the Research Organization Registry with ROR ID 01edhwb26, and on Open Funder Registry with funder ID 100031500. RPD acknowledges funding from the South African Radio Astronomy Observatory (SARAO), which is a facility of the National Research Foundation (NRF), an agency of the Department of Science, Technology and Innovation (DSTI). RPD also acknowledges funding from the South African Research Chairs Initiative of the DSTI/NRF (Grant ID: 77948). We acknowledge the use of the ilifu cloud computing facility – www.ilifu.ac.za, a partnership between the University of Cape Town, the University of the Western Cape, Stellenbosch University, Sol Plaatje University and the Cape Peninsula University of Technology. The ilifu facility is supported by contributions from the Inter-University Institute for Data Intensive Astronomy (IDIA – a partnership between the University of Cape Town, the University of Pretoria and the University of the Western Cape), the Computational Biology division at UCT and the Data Intensive Research Initiative of South Africa (DIRISA). The MeerKAT telescope is operated by the South African Radio Astronomy Observatory, which is a facility of the National Research Foundation, an agency of the Department of Science, Technology and Innovation.

\section{Data Availability}
Code used in this paper is available at \url{https://github.com/FernandoLVentura/vqvae_comparisons}.
The data underlying this article are available in Zenodo, at \url{https://zenodo.org/records/20917414}. 
The datasets were derived from the MeerKAT Galaxy Cluster Legacy Survey \citep{Knowles2022} available at \url{https://doi.org/10.48479/7epd-w356}. 

\section{Conflict of Interest Statement}
Authors declare no conflict of interest.



\bibliographystyle{mnras}
\bibliography{references} 





\bsp	
\label{lastpage}
\end{document}